\documentclass[floatfix,showkeys,amsmath,preprintnumbers,nofootinbib,onecolumn] {revtex4}
\usepackage{graphicx}
\usepackage{colordvi}
\usepackage{amssymb}
\usepackage{latexsym}
    \def\be{\begin{equation}}
    \def\ee{\end{equation}}
    \def\ba{\begin{eqnarray}}
    \def\ea{\end{eqnarray}}
\begin{document}

\title{Astrophysical constraints on primordial black holes in Brans-Dicke theory}

\author{B. Nayak$^a$, A. S. Majumdar$^b$ and L. P. Singh$^a$}
\affiliation{$^a${Department of Physics, Utkal University, Vanivihar,
Bhubaneswar 751004, India}\\
$^b${S. N. Bose National Centre for Basic Sciences, Salt Lake, 
Kolkata 700098, India}\\
E-mail: {bibeka@iopb.res.in, archan@bose.res.in, lambodar\_uu@yahoo.co.in}}

\begin{abstract}
We consider cosmological evolution in Brans-Dicke theory with a population
of primordial black holes. Hawking radiation from the primordial black holes
impacts various astrophysical processes during the evolution of the Universe.
The accretion of radiation by the black holes in the radiation dominated era
may be effective in imparting them a longer lifetime. We present a detailed 
study of how this affects various standard astrophysical constraints coming 
from the evaporation of primordial black holes. We analyze constraints arising
from the present density of the Universe, the present photon
spectrum, the distortion of the cosmic microwave background spectrum and 
also from processes affecting light element abundances after 
nucleosynthesis. We find that the constraints on the initial primordial
black hole mass fractions are tightened with increased accretion efficiency.
\end{abstract}
\maketitle
\tableofcontents
\section{Introduction}
The Standard Model of Cosmology invokes General Theory of Relativity(GTR) 
as the theory of gravity. However, all the confirmative tests 
of GTR have been carried out at low energy. This has led people 
to believe and explore the deviations from GTR at high energy regimes 
like cosmic evolution at very early times. Brans-Dicke (BD) theory of 
gravitation \cite{bd} stands out as one of the most attractive 
alternatives to GTR because it involves minimal extension over GTR through 
introduction of a scalar field $\phi$. In the BD theory the gravitational 
constant becomes function of time and is proportional to the inverse of the 
scalar field $\phi$ which couples to gravity with a coupling parameter $\omega$.
The more general nature of BD theory is evident from the fact that in the 
limit $\omega \to \infty$ it goes over to GTR. Solar system observations 
require $\omega > 10^{4}$ \cite{bertotti}. The ubiquitous nature of BD 
theory is also evident from the fact that it appears in the low energy limit 
of Kaluza-Klein and String theories \cite{am}. Thus, BD theory has been used 
for tackling a number of cosmological problems such as inflation \cite{jl}, 
early and late time behaviour of the Universe \cite{sahoo}, cosmic acceleration 
and structure formation \cite{bermar}, coincidence problem \cite{ns1} 
and problems relating to black holes \cite{ns2}.

It has been pointed out that in the early stage of the Universe Primordial 
Black Holes (PBHs) could be formed due to various mechanisms such as 
inflation \cite{zeld}, initial inhomogeneous conditions \cite{hawk0,bjc1}, 
phase transition \cite{khopol}, bubble collision 
\cite{kss} and decay of cosmic loops \cite{polzem}. 
In the usual formation scenarios, the typical mass of PBHs at the formation 
time could be as large as the 
mass contained in the Hubble volume $M_H$ ranging down to about 
$10^{-4}M_H$ \cite{ams}.
The formation masses of PBHs could be thus small enough to have 
evaporated 
completely by the present epoch due to Hawking evaporation \cite{hawk}. 
However, since the cosmological environment is very hot and dense 
in the radiation-dominated era, it is expected that appreciable absorption of 
the energy-matter from the surroundings could take place. It has
been noticed that such accretion is most effective in altered gravity
scenarios where the PBHs grow due to  
accretion of radiation at a rate smaller than that of the Hubble volume, 
thus providing for enough energy density for the PBHs to accrete causally.
This is responsible for the prolongation of the lifetime of PBHs in 
braneworld models \cite{asm} as well as in scalar-tensor models \cite{mgs}. 

The feasibility of black hole solutions in BD theory was first discussed 
by Hawking \cite{hawking}. Using scalar-tensor gravity theories Barrow and 
Carr \cite{bc} have
studied PBH evaporation during various eras. It has been recently 
observed  \cite{nsm} that in the context of Brans-Dicke 
theory, inclusion of accretion leads to the prolongation of PBH lifetime. 
Once formed, these PBHs will influence later cosmological epochs, leading to a 
number of observational constraints on their allowed abundance. These have 
been extensively investigated in the case of standard cosmology 
\cite{sbs, cgim, bba}. In a recent work, Carr et al. \cite{carretal}
have performed a detailed numerical study of the effect of the emission
of quark and gluons by PBHs on the standard constraints.  The standard
constraints also get altered in different gravitational theories,
as was studied in the context of 
 Brans-Dicke cosmology without accretion
\cite{bc},
and brane-world cosmology \cite{cgl,sns,skns}. The aim of the present paper 
is to 
reanalyse the main constraints in the Brans-Dicke theory with the inclusion
of accretion of radiation in the early Universe. The standard constraint
formalism could get modified due to the change in theory of gravity and also
due to the effect of accretion at early times, as was shown in the case
of braneworld gravity by Clancy {\it et al.} \cite{cgl}. In the present
study we extend our previous work on the evolution 
of PBHs in BD theory including accretion \cite{nsm} to obtain 
several astrophysical constraints following the style of 
Clancy {\it et al.} \cite{cgl}. 

The plan of the paper is as follows. In the next section we first provide
the key expressions related to PBH evaporation, accretion and lifetimes in
the BD scenario, and then discuss the modified constraint formalism in
BD theory that we use subsequently.
In Section III we consider  the observational 
constraints obtained from the present density of the Universe. In Section IV
we discuss the present photon spectrum and constraints following from it. 
The constraints
arising out of the distortion of the Cosmic Microwave Background (CMB) 
spectrum are evaluated in Section V.
Light element abundances 
and photo-disintegration of the deuterium nuclei lead to further constraints
on the initial PBH mass fraction, which are presented in Section VI. 
A summary of our results along with a Table with quantitative estimates
of the various constraints are presented in Section VII.

\section{The constraint formalism in Brans-Dicke theory}

For discussing the constraints that can arise from PBH evaporation,
we label an epoch by cosmic time $t$. PBHs which are not evaporated by 
time $t$ will only contribute to the overall energy density of PBHs. 
As the present observable Universe is nearly flat and, therefore, 
possesses critical density, the PBH mass density can be constrained 
on the ground that it should not overdominate the Universe. 
PBHs evaporate by producing  bursts of evaporation products.
The Hawking radiation from the PBHs which  evaporate well before 
photon decoupling will thermalize with the surroundings, 
boosting the photon-to-baryon ratio \cite{za}. In the case of evaporation after 
photon decoupling, the radiation spectrum is  affected and subsequently 
redshifts in a monotonic manner. Thus, constraints arise from the cosmic 
background radiation at high frequencies \cite{mc, ph, carr}. Further, if the PBHs evaporate
close to the time of photon decoupling, it cannot be fully thermalised 
and will produce distortion in the cosmic microwave background spectrum. 
Generally speaking, at a given epoch, the constraint on various physical 
observables is usually dominated by those PBHs with a lifetime of order of 
the epoch in question. Hence, the observational constraint can be translated
into an upper 
limit on the initial mass fraction of PBHs.

We consider a flat FRW Universe which
is radiation dominated upto a time $t_e$ and  
matter-dominated thereafter.
In the radiation-dominated era \cite{bc}
\be \label{bd1}
a(t) \propto t^{1/2}; ~~~~ G=G_0 \Big(\frac{t_0}{t_e}\Big)^n,
\ee
where $G_0$ denotes the present value of $G$, and $n$ is related to the
Brans-Dicke parameter $w$ by $n=2/(4+3w)$. [In view of the observational bound
on $w$ \cite{bertotti}, we consider the value of $n$ as $n \sim 0.0001$ in our 
subsequent calculations].
The radius and temperature of a PBH is given by
\be \label{bd2}
r=2G_0\Big(\frac{t_0}{t_e}\Big)^n M; ~~~~ T_{BH}=\frac{m_{pl}^2}{8 \pi M} \Big(\frac{t_e}{t_0}\Big)^n.
\ee
Similarly, in the matter-dominated era
\be \label{bd3}
a(t) \propto t^{(2-n)/3}; ~~~~ G(t)=G_0 \Big(\frac{t_0}{t}\Big)^n,
\ee
and
\be \label{bd4}
r=2G_0\Big(\frac{t_0}{t}\Big)^n M ~~ and ~~ T_{BH}=\frac{m_{pl}^2}{8 \pi M} \Big(\frac{t}{t_0}\Big)^n.
\ee

In Brans-Dicke theory the evolution of a PBH \cite{nsm} is described by
\be \label{pbh}
\frac{dM}{dt}=\Big(\frac{dM}{dt}\Big)_{acc} + \Big(\frac{dM}{dt}\Big)_{evap}
\ee
with the accretion and evaporation rates given by
\be \label{acc}
\Big(\frac{dM}{dt}\Big)_{acc}=6fG_0\Big(\frac{t_0}{t_e}\Big)^n \Big(\frac{\dot{a}}{a}\Big)^2 M^2
\ee
and
\be \label{evap}
\Big(\frac{dM}{dt}\Big)_{evap}=-\alpha\Big(\frac{t_e}{t_0}\Big)^{2n} \frac{1}{M^2}
\ee
where $\alpha = \sigma/(256\pi^3G_0^2)$, with $\sigma$ the Stefan-Boltzmann
constant. $f$ is the accretion efficiency which depends 
primarily on two factors: (i) how small the size of the black hole is compared 
to the cosmological horizon, and (ii) how efficient the black hole is in 
absorbing background radiation. The first factor is contingent upon various
formation mechanisms. The second factor could depend, in turn on the thermal
properties of the surrounding radiation, the mean free path of absorbed
particles, as well as the backreaction due the black hole on the cosmological 
metric. Due to several uncertainties in the estimation of the above quantities,
it is usually accepted to consider an accretion efficiency $f$ in the range
between $0$ and $1$.

Accretion makes the mass of PBH to grow a maximum value \cite{nsm} 
\be \label{max}
M_{max}=M(t_c)=\frac{M_i}{1-\frac{3}{2}f}
\ee
where $M_i$ is the initial mass of PBH and $t_c$ is the time at which 
evaporation rate is equal to the accretion rate. Thus $f$ is further restricted 
to lie in between $0$ and $\frac{2}{3}$ in Brans-Dicke theory.
 We assume the standard
mechanism of PBH formation due to the gravitational collapse of density
perturbations at the cosmological horizon scale \cite{bjc1}, which leads to 
$M_i \simeq G^{-1}(t_i)t_i$.

The evaporation time of PBHs which completely evaporate in the radiation 
dominated era, is given by \cite{nsm}
\be \label{tevap}
t_{evap}=t_c\Big[1+ (3\alpha)^{-1}\Big(\frac{t_0}{t_e}\Big)^{2n} \Big(\frac{M(t_c)^3}{t_c}\Big)\Big].
\ee
The PBHs which evaporate in the matter dominated era have their initial mass 
and evaporation time related as \cite{nsm}
\ba \label{mevap}
M_i=\Big{\{}1-\frac{3}{2}f\Big{\}} \times \Big[3\alpha\Big(\frac{t_e}{t_0}\Big)^{2n}t_e\Big{\{}1+(2n+1)^{-1}\times \Big{\{}\Big(\frac{t_{evap}}{t_e}\Big)^{2n+1}-1\Big{\}}\Big{\}}\Big]^{1/3}.
\ea

Constraints on the allowed abundance of PBHs of a certain lifetime are 
formulated as upper bounds on their mass fraction. This mass fraction 
$\alpha_t(M_i)$ is defined as the ratio of the energy density due to PBHs 
of initial mass $M_i$ and the background radiation density, at a 
time $t\geq t_i$, as 
\be \label{alphai}
\alpha_t(M_i)=\frac{\rho_{PBH,M_i}(t)}{\rho_{rad}(t)}.
\ee
The black holes first gain mass through accretion
up to time $t_c$. During this stage $\alpha_t$ scales as $(M a)$. 
Subsequently, they
begin to loose mass very gradually (i.e., the mass stays nearly constant,
and therefore $\alpha_t$ scales as $a$). This stage comes to an end when
$\alpha_t$ reaches the value $\alpha_{evap}$ beyond which the
black hole looses most of its mass in a final burst of evaporation.
It then follows that if initial and final mass fractions denoted by $\alpha_i$ 
and $\alpha_{evap}$ respectively, are related by 
\be \label{e1al}
\alpha_{evap}=\alpha_i \frac{M(t_c)}{M(t_i)} \frac{a(t_{evap})}{a(t_i)}.
\ee
Using equation (\ref{max}), we can write
\be \label{e2al}
\alpha_{evap}=\alpha_i \Big(\frac{1}{1-\frac{3}{2} f}\Big) \frac{a(t_{evap})}{a(t_i)}. 
\ee

The purpose of the following sections is to reconsider observational 
constraints on $\alpha_{evap}$ at different cosmological epochs, trace them 
back to obtain constraints on the initial mass fraction $\alpha_i$ taking
into consideration accretion of radiation in the early universe by the
PBHs in Brans-Dicke theory. It may be mentioned here that the recent
paper by Carr et al. \cite{carretal} has reanalyzed
the standard constraints on primordial black holes by considering the effects
of emission of quarks and gluons and the resultant secondary emission of
photons. It was shown that the effect of secondary photon emission could
alter the standard constraints on PBH fraction by a couple of orders of
magnitude in certain cases, e.g., deuterium constraint, while leaving
the standard constraints more or less unaltered in certain other cases,
e.g., distortion of CMB spectrum. The results obtained by Carr et al. 
\cite{carretal} are
based on detailed numerical analysis. Of course, such a scenario of 
emission would also
impact constraints on Brans-Dicke primordial black holes in more or less
similar ways as they impact PBHs in the standard picture. However, the
lack of analytical results describing the effect of such emission on
the constraint formalism makes it considerably harder to perform  a similar
analysis in the context of an altered gravitational scenario. Since 
the primary aim of the present paper is to study the effect of Brans-Dicke
theory on the constraint formalism, we do not attempt to perform any similar
numerical analysis here. In stead, we make a note of the effects that
emission of secondary photons could have on the different constraints,
based on the results of Ref. \cite{carretal}.

\section{ The present matter density of the Universe}

For a particular value of $f$ in Eq.(\ref{max}) with $t_{evap}=t_0$,
and using Eq.(\ref{mevap}) one gets the 
formation time of PBHs that are evaporating today. Black holes formed later 
are essentially still intact and their density is constrained by the observed 
matter density in the present Universe. Here we are tracking the 
relative densities of PBHs to radiation and we must ensure that, given the 
observed radiation density, this ratio does not imply that the PBH density 
exceeds the observed matter density of about 0.3 of the critical density. 
Phrased in this way, the constraint applies regardless of the presence of a 
cosmological constant and indicates that for any PBHs surviving to the 
present we must have 
\be \label{zero}
\alpha_0(M)<\frac{0.3}{\Omega_{\gamma,0}}.
\ee
The cosmic microwave background corresponds to a photon density of 
$\Omega_{\gamma,0}h^2=2.47 \times 10^{-5}$ (with $h=0.7$) and conservatively, 
we can ignore the cosmic neutrinos.
Thus, for PBHs that are about to evaporate today, $t_{evap}\geq t_0$, we get 
\be \label{aevap}
\alpha_{evap}< \frac{0.3}{\Omega_{\gamma,0}} \approx 6 \times 10^3.
\ee
Using equation(\ref{e2al}), one can find
\be
\alpha_i< 3.43\times 10^{-18} \times \Big(1-\frac{3}{2}f\Big)^{3/2}.
\ee
The constraint on the initial PBH mass fraction in the tandard cosmology is 
obtained as \cite{cnpsz}
\be \label{alphai1}
\alpha_i< 10^{-18}.
\ee

\section{The present photon spectrum}
If PBHs evaporate between the time of photon decoupling($t_{dec}$) and the 
present day, their radiation spectra will not be appreciably influenced by the 
background Universe, apart from being redshifted. Thus the spectra could 
constitute a fraction of the cosmic background radiation. 
The number of particles of a certain species j, emitted in 4-dimensional 
spacetime by a black hole of temperature $T_{BH}$, in a time interval $dt$ 
and with momentum in the interval (k, k+dk) is 
\be \label{spe1}
dN_j=\sigma_j(k) \frac{dt}{exp(\frac{\omega}{T_{BH}})\pm 1} \frac{d^3k}{(2\pi)^3}
\ee
with $\omega^2=k^2+m^2$, where  $\sigma_j(k)$ is the
emission cross-section for species $j$ with momentum $k$ and $m$ is the mass 
of the particle.

For emission of photons with energy in the interval (E, E+dE) above equation becomes
\be \label{spe2}
dN=\frac{\sigma(E)}{2\pi^2} \frac{E^2}{exp(\frac{E}{T_{BH}})-1} dE dt.
\ee
Thus the spectral photon number emitted by a small black hole with lifetime $t_{evap}$ is obtained from
\ba \label{spe3}
\frac{dN}{dE}=\int_{0}^{t_{evap}} {\frac{\sigma(E)}{2\pi^2} \frac{E^2}{exp(\frac{E}{T_{BH}})-1}} \,dt.
\ea
But, in the high frequency limit $E>>T_{BH}$, all cross section reduce to the same value \cite{cgl} as 
\be  \label{sig}
\sigma=4\pi r_0^2=\frac{1}{4 \pi T_{BH}^2}.
\ee
Now, the spectral number becomes
\ba \label{spe4}
\frac{dN}{dE}=\frac{1}{8\pi^3} \int_{0}^{t_{evap}} {\frac{(\frac{E}{T_{BH}})^2}{exp(\frac{E}{T_{BH}})-1}} \,dt.
\ea 
In the high energy limit, the above integral yields the spectrum \cite{cgl}
\ba \label{spe6}
\frac{dN}{dE}=\frac{24.886}{4096\pi^6} \alpha^{-1} \Big(\frac{t_e}{t_0}\Big)^n m_{pl}^6 E^{-3}.
\ea
(where $\alpha$ has been defined below Eq.(\ref{evap}).)
The spectrum declines as $E^{-3}$ like the standard spectrum.

Now, consider black holes evaporating at a time $t_{evap} \geq t_{dec}$. 
We make the approximation that all the energy gets released instantly, 
but take the spectrum into account. The black hole mass fraction just 
before evaporation is given by 
\be \label{alev}
\alpha_{evap}=\alpha_i \frac{M(t_c)}{M_i} \frac{a(t_{evap})}{a(t_i)},
\ee
while the number density (which scales as $a^{-3}$) is
\be \label{eta1}
n_{evap}(M_i)=n_{evap}(M_c) \frac{a(t_c)^3}{a(t_i)^3},
\ee
where
\be \label{eta2}
n_{evap}(M_c)=\alpha_{evap} \frac{\rho_{rad}(t_{evap})}{M(t_c)}
\ee
leading to
\ba \label{eta3}
n_{PBH}(t_{evap})=\alpha_{evap} \frac{\rho_{rad}(t_{evap})}{M(t_c)}\Big(1-\frac{3}{2}f\Big)^{-3/2}.
\ea

The energy density in photons of energy $E$ emitted between $t_{evap}$ and $t_{evap}+dt_{evap}$ is \cite{cgl}
\be \label{en1}
dU_{tev}(E)=n_{PBH}(t_{evap}) E^2 \frac{dN}{dE}(t_{evap}) \frac{dt_{evap}}{t_{evap}}.
\ee
We require the present total energy density in Hawking photons at a certain energy scale $E_0$, denoted as $U_0(E_0)$. 

The photon energy $E$ emitted at time $t$ undergoes 
redshift due to expansion of the Universe. 
So the photons emitted at time $t_{evap}$ with energy $E$ is related 
with presently observed energy as
\be \label{en2}
E(E_0)=E_0 \frac{a(t_0)}{a(t_{evap})}.
\ee
Since
\be \label{rho}
\rho_{rad}(t_{evap})=\rho_{rad}(t_0) \frac{a(t_0)^4}{a(t_{evap})^4},
\ee
one gets
\ba \label{ue1}
U_0(E_0)=\int_{t_{dec}}^{t_0} \,dU_0(E_0) =\int_{t_{dec}}^{t_0}{\frac{a(t_{evap})^4}{a(t_0)^4}} \,dU_{tev}[E(E_0)].
\ea
Substituting the relevant expressions and using Eq.(\ref{mevap}), we get
\ba \label{ue3}
U_0(E_0)=m_{pl} \Big(\frac{t_0}{t_e}\Big)^{(2-n)/3} (t_e)^{\frac{1}{2}+n} \Big(\frac{t_0}{t_e}\Big)^{\frac{n}{2}} (3\alpha)^{-1/2} \rho_{rad}(t_0) E_0^2 t_0^{(2-n)/3}(2n+1)^{\frac{1}{2}} \nonumber \\ \int {{\alpha_i}\Big(1-\frac{3}{2}f\Big)^{-3}(t_{evap})^{-(\frac{1}{2}+n)}} (t_{evap})^{-(2-n)/3} \frac{dN}{dE(E_0)}(t_{evap}) \frac{dt_{evap}}{t_{evap}}. 
\ea

The number spectrum of a black hole of initial temperature $T_{BH}$ peaks at 
an energy $E=bT_{BH}$ with $b\approx5$ \cite{page} in the standard treatment. 
In the BD case the $b$-value remains unchanged since in its determination from 
the power spectrum of the emitted particles,  $G$ which in this case is 
$\sim \phi^{-1}$, gets cancelled. Therefore, unless $\alpha_i$ is sharply 
peaked at particular initial epochs, the main contribution to the integral 
in Eq.(\ref{ue3}) is obtained when $E(E_0)=bT_{BH}(t_{evap})$, i.e., from PBHs 
evaporating at $t_{evap}=t_{main}$, where
\be \label{main}
t_{main} \approx \Big(\frac{E_0}{bT_{BH}(t_0)}\Big)^{3/1-3n} t_0.
\ee
The contribution from PBHs evaporating earlier will come from the high 
frequency end of their spectrum, while PBHs evaporating at later times will 
contribute radiation that originated in the low frequency end. 
Using $x_i=b\approx5$, the number  spectrum equation (\ref{spe4}) becomes
\ba \label{spe7}
\frac{dN}{dE(E_0)}(t_{evap})=\frac{14.3}{4096\pi^6} \frac{m_{pl}^6}{\alpha} \Big(\frac{t_e}{t_0}\Big)^n E_0^{-3} \Big(\frac{a(t_{evap})}{a(t_0)}\Big)^3.
\ea
Using this equation in equation(\ref{ue3}), we estimate the total energy 
density at energy $E_0$ as
\ba \label{ue4}
U_0(E_0)=\frac{14.3}{4096\pi^6} \frac{m_{pl}^6}{\alpha} \Big(\frac{t_e}{t_0}\Big)^n  m_{pl} \Big(\frac{t_0}{t_e}\Big)^{(4+n)/6} (t_e)^{\frac{1}{2}+n} (3\alpha)^{-1/2}\rho_{rad}(t_0) E_0^{-1} t_0^{(2-n)/3}(2n+1)^{\frac{1}{2}} \nonumber \\ \int_{t_{dec}}^{t_{main}} {{\alpha_i}(1-\frac{3}{2}f)^{-3}(t_{evap})^{-(1+10n)/6}} \, dt_{evap},
\ea
which gives 
\be \label{ue5}
U_0(E_0) \propto E_0^{(3-16n)/(2-6n)}.
\ee
In the standard spectrum $U_0(E_0)$ varies as $E_0^{3/2}$.

Radiation at lower frequencies will originate from the low frequency 
ends of the instantaneous spectra, with the dominant contribution coming from 
PBHs evaporating around $t_{dec}$. Its intensity can generically be neglected 
as compared to the main frequency range \cite{cgl}. For energies 
$E_0 > bT_{BH}(t_0)$, 
the dominant part comes from the high frequency tail of PBHs evaporating today. 
The number spectrum equation in the high energy limit  is used to obtain
\ba \label{ue6}
U_0(E_0)=6.16 \times 10^{54} E_0^{-1} \alpha_i \Big(1-\frac{3}{2}f\Big)^{-3} ~~ keV cm^{-3} s^{-2} keV^{-1}.
\ea

The spectral surface brightness $I(E_0)$, an observational quantity, is related
to  the integrated 
energy density $U_0(E_0)$ by \cite{sns}
\be \label{ie1}
I(E_0)=\frac{c}{4\pi} \frac{U_0(E_0)}{E_0}.
\ee
The overall peak in the present spectrum is at $E_{peak}=bT_{BH}(t_0)$ . So equation(\ref{ie1}) gives
\ba \label{ie2}
I(E_{peak})=2.08 \times 10^{20} \Big(1-\frac{3}{2}f\Big)^{-1} \alpha_i ~~  keV cm^{-2}s^{-1}sr^{-1}keV^{-1}.
\ea
Considering a value of $E_{peak} \approx 100$ Mev \cite{ph}, which is not
too far from the present observational range \cite{smr} of $\gamma$-rays, 
we have 
\be \label{ie3}
I_{obs}=1.11 \times 10^{-5} ~~ keV cm^{-2}s^{-1}sr^{-1}keV^{-1}.
\ee
The constraint $I(E_{peak})<I_{obs}$ then results in an upper limit on the 
initial mass fraction as
\be \label{alphai2}
\alpha_i<5.34 \times 10^{-26} \times \Big(1-\frac{3}{2}f\Big).
\ee
For comparison, the corresponding constraint in the standard case is obtained 
from the gamma-ray background at $E_{peak}=100MeV$ and reads $\alpha_i<10^{-27}$ \cite{cgim, ph, carr}.  It may be noted here that the recent work of Carr et al. 
\cite{carretal} confirms earlier results \cite{macweb} that the
spectrum of secondary photons is peaked at $E \approx 68MeV$, independent of
the PBH temperature. Our constraint on the PBH mass fraction originates
from considering the peak value $E \approx 100 MeV$ for primary photons.
As shown by Carr et al. \cite{carretal}, secondary photon emission may 
dominate when the
PBH mass falls below the QCD scale, in which case the constraints may
be altered at most by an order of magnitude. Similar considerations would
apply to the Brans-Dicke case, but the exact magnitude of the constraint
would require numerical analysis to be evaluated.

\section{Distortion of the cosmic microwave background spectrum}
Hawking radiation emitted at redshifts $z \leq 2 \times 10^{6}$ or 
$t \geq 4 \times 10^{-10}t_0$ can not be fully thermalised and will disturb 
the Cosmic Microwave Background spectrum. The corresponding modification of the 
Planck spectrum is described by the chemical potential $\mu$ \cite{szm}, which 
is related with injected energy density as
\be \label{rrho}
\frac{\rho_{evap}}{\rho_{rad}(t)}=0.71 \mu,
\ee
where $\rho_{evap}$ is the energy density injected by evaporating PBHs. 
Observational results \cite{fixsen} suggest an upper limit on $\mu$ given by
\be \label{mu}
\mu < 9 \times 10^{-5}.
\ee
Assuming that about a half of the energy is emitted in the form of particles capable to disturb the CMB spectrum, one can write
\be \label{fmu}
\frac{1}{2}\alpha_{evap}=0.71 \mu.
\ee
Using equation (\ref{mu}), we get
\be \label{haev}
\alpha_{evap}<1.28 \times 10^{-4},
\ee
which leads to the initial PBH mass fraction constraint as
\be \label{alphai3}
\alpha_i<1.28 \times 10^{-21} \times\Big(1-\frac{3}{2}f\Big)^{3/2}.
\ee
In the standard cosmology the corresponding constraint is \cite{nasel}  
\be \label{alph}
\alpha_i<10^{-21}.
\ee

In a recent work it was shown by Tashiro and Sugiyama \cite{tasug}, 
that secondary photon 
emissions leading to non-zero
chemical potential for photons could impact the CMB spectrum. However, it
follows from the analysis of Carr et al. \cite{carretal}, that the constraint 
on the PBH mass 
fraction stays around the value $10^{-21}$.

\section{Nucleosynthesis constraints}
Standard big-bang primordial 
nucleosynthesis is one of the most well-understood processes in the early
Universe. Therefore, this era is an important benchmark to look for
effects due to the interactions of particles emitted by PBHs \cite{carr}. 
Several detailed investigations of PBHs in standard cosmology have computed
the predicted changes in the density of light elements \cite{vdn,ky}. Existing 
observational limits on the light element abundances have then been used 
to put constraints on the size of such modifications, which in turn
lead to  constraints on the numbers of PBHs that could evaporate both 
during and after nucleosynthesis.  In the standard scenario, the PBHs which 
evaporate during nucleosynthesis 
with 
$t_{evap}$  between $1$s to $400$s, have initial masses in between 
$ 10^9$g and $10^{10}$g. This remains nearly the same in Brans-Dicke theory 
where the 
initial mass varies between $3.1 \times 10^9 \times (1-\frac{3}{2}f) $g 
and $2.28 \times 10^{10} \times (1-\frac{3}{2}f) $g. Following again the 
analysis of Clancy,
Guedens and Liddle \cite{cgl} (performed in the context of braneworld
black holes) here we examine two nucleosynthesis 
constraints in the context of the Brans-Dicke theory, namely the constraint 
on the increase in production of helium-4 due to the injection of PBH 
hardons \cite{ky, zskc} and the constraint on the destruction of primordial 
deuterium by PBH photons \cite{ky, lindley}. Note here that the 
nucleosynthesis constraints are the ones that could be most affected by
taking into account quark and gluon emission by PBHs. Inter-conversion
between protons and neutrons due to emitted mesons and anti-nucleons
increases the n/p freeze-out ratio as well as the final He-4 abundance,
as shown by Carr et al. \cite{carretal} in their recent work. Our main aim here is to get a 
reasonable estimate of how such constraints are modified in BD theory.
\subsection{The Helium abundance constraint}
The total number density of emitted particles from the complete evaporation 
of PBHs of some initial mass may be expressed as 
\be \label{nem}
N_{em}=\frac{\rho_{PBH}}{<E_{em}>},
\ee
where $<E_{em}>$ is the average energy of the emitted particles.
The ratio of the energy density in PBHs at evaporation to the background 
radiation energy density is therefore
\be \label{fpro}             
\frac{\rho_{PBH}}{\rho_{rad}}=\alpha_{evap}=\frac{<E_{em}>N_{em}}{<E_{rad}>N_{rad}},
\ee
where $<E_{rad}>$ and $N_{rad}$ are the average energy and number density of 
the particles comprising the background cosmological radiation fluid.
The ratio of the average energies can be approximated by 
the ratio of the PBH temperature at the onset of evaporation to the background 
temperature at evaporation, i.e.,
\be \label{eem}
\frac{<E_{em}>}{<E_{rad}>}=\frac{T_{BH}}{T_{evap}}.
\ee
Using equation (2)--(8) and applying the standard cosmological temperature-time
relation \cite{kt}
\be \label{ttt}
t=0.301 g_*^{-1} \frac{m_{pl}}{T^2}
\ee
(where $g_*$ is a constant having value $10.78$ ),
one gets
\ba \label{frac}
\frac{<E_{rad}>}{<E_{em}>}=\frac{T_{evap}}{T_{BH}}=7.28 \times \Big(1-\frac{3}{2}f\Big)  \Big(\frac{M(t_c)}{m_{pl}}\Big)^{-1/2}.
\ea
The total emitted number density at  time $t_{evap}$ during nucleosynthesis 
thus becomes
\be \label{fnem}
N_{em}=7.28 \Big(1-\frac{3}{2}f\Big) \alpha_{evap}  \Big(\frac{M(t_c)}{m_{pl}}\Big)^{-1/2}
N_{rad}.
\ee
Using observational estimations, Clancy {\it et. al}.\cite{cgl} found that 
\be \label{onem}
N_{em}<\frac{2.8}{100F} n_b,
\ee
where $F$ is the fraction of the total particles emitted by PBH having value 
$F \leq 0.2$.
Comparing above two equations, one can get
\ba \label{alpe}
\alpha_{evap}<\frac{0.38}{F} \times 10^{-2} \Big(1-\frac{3}{2}f\Big)^{-1}  \Big(\frac{M(t_c)}{m_{pl}}\Big)^{1/2} \eta_{evap},
\ea
where $\eta_{evap}=\frac{n_b}{N_{rad}}$ is the baryon to photon ratio at 
evaporation. Assuming that $\eta$ is fixed from evaporation up to 
present times, i.e., $\eta_{evap}=\eta_0$ and using the relation 
$\eta_0 \approx 2.8 \times 10^{-8} \Omega_bh^2$ with $\Omega_bh^2 \approx 0.02$, Eq.(\ref{alpe}) becomes
\be \label{alphe}
\alpha_{evap}<1.064 \times 10^{-11} \times \Big(1-\frac{3}{2}f\Big)^{-1} \Big(\frac{M(t_c)}{m_{pl}}\Big)^{1/2}.
\ee
The bounds imposed at nucleosynthesis can be converted into bounds on the 
initial PBH mass fraction for PBHs that evaporate at $t_{evap}=400$s:
\be \label{alphai3}
\alpha_i<8 \times 10^{-20} \times  \Big(\frac{M_i}{10^9g}\Big)^{1/2}.
\ee
One may compare the constraint with that of  standard cosmology \cite{ky, zskc} 
\be \label{salpe}
\alpha_i<3 \times 10^{-18}  \Big(\frac{M_i}{10^9g}\Big)^{1/2}.
\ee
Equation(\ref{alphai3}) further leads to 
\be \label{malphai3}
\alpha_i<3.82 \times 10^{-19} \times  \Big(1-\frac{3}{2} f\Big)^{1/2}.
\ee

It turns out from the analysis by Carr et al. \cite{carretal} that by
considering the effects of quark and gluon emission by the PBHs, an earlier
constraint \cite{ky} 
on the PBH mass fraction is actually weakened in the relevant mass range
that we are also considering for our analysis in the Brans-Dicke case.

\subsection{Deuterium photodisintegration constraint}

The high-energy particles emitted by evaporating PBHs both during and after 
nucleosynthesis can be sufficiently energetic to disrupt primordial nuclei. 
One important reaction of this type is photo-disintegration which entails
the destruction of primordial nuclei by high-energy PBH photons. 
Of all the primordial nuclei, deuterium is the most susceptible to 
photon-disintegration.
If $\Delta M$ is the PBH mass evaporated between times $t_1$ and $t_2$ during 
which deuterons are destroyed and $M_b$ is the baryonic mass, then 
\cite{lindley}
\be \label{ame}
\frac{\Delta M}{M_b} \leq \frac{\epsilon}{f_{\gamma} \beta} \frac{E_*}{m_p},
\ee
where $f_{\gamma}$ is the fraction of mass that decays into photons, $\epsilon$ is the depletion factor, $E_*$ and $\beta$ are constants.
$t_1$ and $t_2$ are taken to be the end of nucleosynthesis and the onset 
of recombination respectively. We estimate  $\Delta M$ by the PBH mass 
evaporated shortly after nucleosynthesis. It is usually justified to take 
\cite{lindley} 
\be \label{dme}
\frac{\Delta M}{M_b}=\Big[\frac{\rho_{PBH}}{\rho_b}\Big]_{t_{evap}}
\ee
with $t_{evap}$ some time after nucleosynthesis, either when a narrow mass 
range of PBHs evaporates, or straight after nucleosynthesis for an extended 
mass spectrum. Equation (\ref{ame}) then leads to 
\be \label{eps}
\alpha_{evap} \leq \Big[\frac{\rho_{b}}{\rho_{rad}}\Big]_{t_{evap}}\frac{\epsilon}{f_{\gamma} \beta} \frac{E_*}{m_p}.
\ee
Since $\rho_b \propto a^{-3}$ and $\rho_{rad} \propto a^{-4}$, it follows that
\ba \label{rhob}
\Big[\frac{\rho_{b}}{\rho_{rad}}\Big]_{t_{evap}}=\frac{a(t_{evap})}{a(t_e)}\Big[\frac{\rho_{b}}{\rho_{rad}}\Big]_{t_e}=2 \frac{a(t)}{a(t_e)} \Omega_b(t_e)
\ea
as $\rho_{rad}=\rho_{tot}/2 \approx \rho_c/2$ at the time of matter radiation equality $t_e=10^{11}$s. The baryon density parameter at equality is related to the present one, as the matter density parameter at equality is given by $\Omega_m(t_e) \approx {1/2}$.
\ba \label{omegat}
\Omega(t_e) =\frac{\Omega_m(t_e)}{\Omega_m(t_0)} \Omega_b(t_0) \approx \frac{1}{2} \frac{\Omega_b(t_0)}{\Omega_m(t_0)}.
\ea
Using equation (\ref{omegat}) in equation (\ref{eps}), one gets 
\ba \label{estar}
\alpha_{evap} \leq \frac{\epsilon}{f_{\gamma} \beta} \frac{E_*}{m_p} \Big(\frac{t_{evap}}{t_e}\Big)^{1/2} \frac{\Omega_b(t_0)}{\Omega_m(t_0)}.
\ea
Using numerical values of constants \cite{cgl} $f_{\gamma}=0.1$,  $E_*=10^{-1}$Gev,  $\beta=1$,  $\epsilon \sim 1$ and taking the ratio of the present baryonic density to the total matter density as $0.1$, we get
\be \label{ceva}
\alpha_{evap} \leq 1.066 \times 10^{-28} \Big(\frac{t_{evap}}{t_{pl}}\Big)^{1/2}.
\ee
For $t_{evap}=400$ sec, the constraint becomes
\be \label{evaf}
\alpha_{evap} \leq 6.74 \times 10^{-6},
\ee
which gives
\be \label{alphai4}
\alpha_i \leq 5.1 \times 10^{-21}\times\Big(1-\frac{3}{2}f\Big)^{3/2}.
\ee
In standard case this constraint is \cite{ky, lindley}  
\be \label{als}
\alpha_i \leq 10^{-21}.
\ee

The recent analysis by Carr et al. \cite{carretal} reveals that the 
standard deuterium constraint \cite{ky}
is tightened by two orders of magnitude, since hadrodissociation of helium
due to injected nuclei produces more deuterium. The same is expected to be 
true for the Brans-Dicke
scenario, but, as in the earlier cases, only detailed numerical simulations
could reveal the actual quantitative changes on the constraints.



\section{Summary and Conclusions}

\begin{table*}
\caption{The variation of the upper bound of the initial PBH mass fraction($\alpha_i$) with the accretion efficiency $f$ for different cases is shown in the table}
\begin{ruledtabular}
\begin{tabular}[c]{ccccc}
Cause of the Constraint  & $f=0$ & $f=0.25$ & $f=0.45$ & $f=0.65$ \\
\hline
Present Density & $3.43\times10^{-18}$ & $1.69\times10^{-18}$ & $0.62\times10^{-18}$  &  $0.01\times10^{-18}$\\
\hline
Photon Spectrum & $5.34\times10^{-26}$ & $3.34\times10^{-26}$ & $1.73\times10^{-26}$ & $1.33\times10^{-27}$\\
\hline
Distortion of CMB & $1.28\times10^{-21}$ & $0.63\times10^{-21}$ & $0.23\times10^{-21}$ & $0.05\times10^{-22}$\\
\hline
Helium abundance & $3.82\times10^{-19}$ & $3.01\times10^{-19}$ & $2.18\times10^{-19}$ & $0.60\times10^{-19}$\\
\hline
Deuterium abundance & $5.10\times10^{-21}$ & $2.52\times10^{-21}$ & $0.94\times10^{-21}$ & $0.02\times10^{-21}$\\
\end{tabular}
\end{ruledtabular}
\end{table*}

In this paper we have analyzed several astrophysical constraints on
the initial mass fraction of primordial black holes that could grow
due to accretion of radiation in the early Universe in Brans-Dicke
cosmology. It is well-known that PBHs loose mass due to Hawking 
evaporation with their lifetime depending upon their initial masses. 
However, the  accretion of radiation could be an efficient process
in the BD formalism, depending upon the accretion 
efficiency, leading to the growth of mass and thereby  enhancing their 
lifetime \cite{mgs,nsm}. Such PBHs, once formed, will influence the later 
cosmological epochs through their total density, and also through the
products of their evaporation. Since the standard cosmological scenario
is based on a sound observational footing, at least from the era of
nucleosynthesis onwards, any modification due to surviving PBHs at various
eras is tightly constrained by various observational results, such as the 
photon spectrum, CMB radiation, light element abundances, etc. Such 
constraints, in turn, can be translated into constraints on the initial PBH 
mass fraction. This translation is contingent on the particular cosmological
evolution, and is thus sensitive to the theory of gravity considered, as
has been shown earlier in the context of BD gravity without accretion
\cite{bc}, and also in the context of braneworld gravity 
\cite{cgl}. Moreover, inclusion of the effect of accretion also impacts
upon the constraints, as seen earlier in the context of braneworld
gravity \cite{cgl}, and also as shown by us in the present work.

The summary of our results are presented in Table 1. Here we enlist the 
upper bounds on the initial mass fraction of PBHs that are allowed by taking
into considerations various observational features as listed. These constraints
obviously depend upon the accretion efficiency, as seen from the displayed
numbers. The initial constraints are usually the most severe for those black 
holes whose lifetimes are comparable with the cosmic time of the epoch at 
which the observational constraint is imposed, and our displayed results
correspond to such cases. We consider three different values of the accretion
efficiency, with $f \approx 2/3$ corresponding to the maximum allowed
in BD theory. The $f=0$ case corresponds to the BD formalism without
accretion. It is seen that in all cases the inclusion of 
accretion strengthens the constraint on the upper bound of the
initial PBH mass fraction. 
 
Since the present observable Universe is flat, the mass density of presently 
surviving PBHs should not exceed that of dark matter which is about $0.3$ 
times the critical density. This imposes a constraint on initial PBH mass 
fraction which is of order $10^{-18}$ for lower accretion efficiencies and 
grows by two orders for higher accretion efficiencies. The standard cosmology 
constraint is of order $10^{-18}$. PBHs which evaporate between photon 
decoupling and the present age leave behind a spectrum that peaks at a 
temperature of the order of the black hole temperature at the onset of 
evaporation 
of PBHs with $t_{evap} \approx t_0$. In this case the bound on the
initial PBH mass 
fraction is comparable with standard cosmology for higher accretion 
efficiencies and grows weaker by one order for lower accretion efficiencies. 
If evaporation products are released around the Sunyayev-Zel'dovich time 
$t_{SZ} \approx 4 \times 10^{-10}t_0$, they will fail to fully thermalize 
the background radiation. 
This time, however, is sufficiently early for the excess energy to distort the 
background blackbody spectrum. Limits on the allowed distortion of the CMB 
spectrum then imply limits on PBH mass fraction. For lower accretion 
efficiencies this constraint is of the same order as in standard cosmology, but 
increases more than three orders for maximally efficient accretion. If there 
were a population of PBHs evaporating during or after the era of 
nucleosynthesis, 
this could have led to significant change in the final light element 
abundances. Considering the Helium abundance as an example, we have found 
that the constraint on the initial PBH mass fraction once again varies as it 
started from a value of nearly one order less compared to standard cosmology
and grows closer to it as accretion efficiency increases. Considering 
photon-disintegration and change in deuterium abundance, we find that the 
initial constraint is nearly of the same order as the standard value for 
low accretion 
efficiencies, increasing by nearly two orders for higher values. 
Comparing the constraints due to the different observational features
considered in this work, we note that the photon spectrum imposes
the most stringent limits on the initial PBH mass fraction.

We conclude by emphasizing that Brans-Dicke cosmology which provides a 
viable alternative to the standard scenario, imposes upper bounds on the
allowed initial mass fraction of primordial black holes, that are modified
compared to standard cosmology. Depending upon the particular observed
physical process used to impose the constraints, these upper bounds 
in BD gravity could either be strengthened or weakened compared to the case of
standard gravity \cite{bc}.  It needs to be mentioned here that
some of the constraints of standard gravity could themselves be modified
by considering effects of quark and gluon emission and the resultant emission
of secondary photons by PBHs, as discussed in the recent work by Carr et
al. \cite{carretal}. However, as shown in the present paper the inclusion of the effect of accretion tightens the constraints in all cases since PBHs in
BD gravity could start with a lower value of initial mass, and subsequently
grow in size sufficiently \cite{nsm} to impact the observational
features in future eras. Finally, it remains to be seen how additional
effects such as considering PBHs in further general scalar-tensor models of 
gravity such as in \cite{mgs}, or taking into account the effects of
backreaction of the PBHs on cosmological evolution, could modify the
observational constraints on the initial mass spectrum.

\section*{Acknowledgements}
B. Nayak would like to thank the Council of Scientific and Industrial Research, Government of India, for the award of SRF, F.No. $09/173(0125)/2007-EMR-I$. 



\begin{thebibliography}{99}
\bibitem{bd} 
C. Brans and R. H. Dicke, 
\textit{Mach's Principle and a Relativistic Theory of Gravitation}, 
Phys. Rev. D \textbf{124}, 925 (1961).

\bibitem{bertotti}
B. Bertotti, L. Iess, and P. Tortora, 
\textit{ A test of general relativity using radio links with the Cassini spacecraft}, 
Nature (London) \textbf{425}, 374 (2003).


\bibitem{am} 
A. S. Majumdar and S. K. Sethi, 
\textit{ Extended inflation from Kaluza-Klein theories}, 
Phys. Rev D \textbf{46}, 5315 (1992) ;
A. S. Majumdar, T. R. Seshadri and S. K. Sethi, 
\textit{ Stable compactification and inflation from higher dimensional Brans-Dicke theory}, 
Phys. Lett. B \textbf{312}, 67 (1993) ;
A. S. Majumdar, 
\textit{Constraints on higher dimensional models for viable extended inflation},
Phys. Rev. D \textbf{55}, 6092 (1997) [gr-qc/9703070].


\bibitem{jl}
C. Mathiazhagan and V. B. Johri, 
\textit{ An inflationary universe in Brans-Dicke theory: a hopeful sign of theoretical estimation of the gravitational constant}, 
Class. Quantum Grav \textbf{1}, L29 (1984) ; 
D. La and P. J. Steinhardt, 
\textit{Extended Inflationary Cosmology}, 
Phys. Rev. Lett. \textbf{62}, 376 (1989).


\bibitem{sahoo} 
B. K. Sahoo and L. P. Singh, 
\textit{ Time dependence of Brans-Dicke parameter omega for an expanding universe},  
Mod. Phys. Lett. A \textbf{17}, 2409 (2002) [gr-qc/0210004] ; 
\textit{ Cosmic evolution in generalised Brans-Dicke theory}, 
Mod. Phys. Lett. A \textbf{18}, 2725 (2003) [gr-qc/0211038]. 


\bibitem{bermar} 
O. Bertolami and P. J. Martins, 
\textit{Nonminimal coupling and quintessence}, 
Phys. Rev. D \textbf{61}, 064007 (2000) [gr-qc/9910056].


\bibitem{ns1} 
B. Nayak and L. P. Singh, 
\textit{Present Acceleration of the Universe, Holographic Dark Energy and Brans-Dicke Theory},
Mod. Phys. Lett. A \textbf{24}, 1785 (2009) [arXiv:0803.2930].

\bibitem{ns2} 
B. Nayak and L. P. Singh, 
\textit{ Brans-Dicke Theory and PBH in Early Matter-Dominated Era}, 
arXiv:0905.3657.

\bibitem{zeld}
B. J. Carr, J. H. Gilbert and J. E. Lidsey,
\textit{Black hole relics and inflation: Limits on blue perturbation spectra},
Phys. Rev. D \textbf{50}, 4853 (1994) [astro-ph/9405027] ;
P. Ivanov, P. Naselsky and I. Novikov, 
\textit{Inflation and primordial black holes as dark matter}, 
Phys. Rev. D \textbf{50}, 7173 (1994) ; 
J. Garcia-Bellido, A. D. Linde and D. Wands, 
\textit{Density perturbations and black hole formation in hybrid inflation}, 
Phys. Rev. D \textbf{54}, 6040 (1996) [astro-ph/9605094] ;  
J. Yokoyama, 
\textit{Formation of MACHO-primordial black holes in inflationary cosmology}, 
Astron. Astrophys. \textbf{318}, 673 (1997) [astro-ph/9509027] ;  
J. Yokoyama,
\textit{Chaotic new inflation and formation of primordial black holes},
Phys. Rev. D \textbf{58}, 083510 (1998) [astro-ph/9802357] ;
M. Kawasaki and T. Yanagida, 
\textit{Primordial black hole formation in supergravity}, 
Phys. Rev. D \textbf{59}, 043512 (1999) [hep-ph/9807544] ;  
T. Kawaguchi, M. Kawasaki, T. Takayama, M. Yamaguchi, J. Yokoyama, 
\textit{Formation of intermediate-mass black holes as primordial black holes in the inflationary cosmology with running spectral index},  
Mon. Not. Roy. Astron. Soc. \textbf{388}, 1426 (2008) [arXiv:0711.3886] ; 
L. Alabidi and K. Kohri, 
\textit{Generating primordial black holes via hilltop-type inflation models}, 
Phys. Rev. D \textbf{80}, 063511 (2009) [arXiv:0906.1398].

\bibitem{hawk0}
 S. W. Hawking,
\textit{Gravitationally collapsed objects of very low mass}, 
Mon. Not. R. Astron. Soc. \textbf{152}, 75 (1971).

\bibitem{bjc1}
 B. J. Carr,
\textit{ The Primordial black hole mass spectrum}, 
Astrophys. J. \textbf{205}, 1 (1975).


\bibitem{khopol}
 M.Y.Khlopov and A.Polnarev,
\textit{ Primordial Black Holes As A Cosmological Test Of Grand Unification}, 
Phys. Lett. B \textbf{97}, 383 (1980); 
J. C. Niemeyer and K. Jedamzik,
\textit{Near-Critical Gravitational Collapse and the Initial Mass Function of Primordial Black Holes}, 
Phys. Rev. Lett. \textbf{80}, 5481 (1998) [astro-ph/9709072] ; 
J. C. Niemeyer and K. Jedamzik,
\textit{Dynamics of primordial black hole formation}, 
Phys. Rev. D \textbf{59}, 124013 (1999) [astro-ph/9901292] ; 
K. Jedamzik and J. C. Niemeyer,
\textit{Primordial black hole formation during first-order phase transitions}, 
Phys. Rev. D \textbf{59}, 124014 (1999) [astro-ph/9901293] ;
S. G. Rubin, M. Y. Khlopov and A. S. Sakharov, 
\textit{ Primordial black holes from nonequilibrium second order phase transition}, 
Grav. Cosmol. \textbf{S6}, 51 (2000) [hep-ph/0005271] ; 
K. Nozari,
\textit{A possible mechanism for production of primordial black holes in early universe}, 
 Astropart. Phys. \textbf{27}, 169 (2007) [arXiv:hep-th/0701274] ; 
I. Musco, J. C. Miller and A. G. Polnarev, 
\textit{Primordial black hole formation in the radiative era: investigation of the critical nature of the collapse}, 
Class. Quant. Grav. \textbf{26}, 235001 (2009) [arXiv:0811.1452].


\bibitem{kss}
 H.Kodama, M.Sasaki and K.Sato,
\textit{Abundance of Primordial Holes Produced by Cosmological First-Order Phase Transition}, 
Prog. Theor. Phys. \textbf{68}, 1979 (1982).

\bibitem{polzem}
 A. Polnarev and R. Zembowicz,
\textit{Formation of primordial black holes by cosmic strings}, 
Phys. Rev. D \textbf{43}, 1106 (1991); 
J. C. Hildago and A. G. Polnarev,
\textit{Probability of primordial black hole formation and its dependence on the radial profile of initial configurations}, 
Phys. Rev. D \textbf{79}, 044006 (2009) [arXiv:0806.2752].


\bibitem{ams}
I. Hawke and J. M. Stewart,
\textit{The dynamics of primordial black-hole formation},  
Class. Quant. Grav. \textbf{19}, 3687 {2002}.

\bibitem{hawk}
 S. W. Hawking, 
\textit{ Particle Creation by Black Holes}, 
Commun. Math. Phys. \textbf{43}, 199 (1975).  
 
\bibitem{asm}
 A. S. Majumdar, 
\textit{ Domination of black hole accretion in brane cosmology}, 
Phys. Rev. Lett. \textbf{90}, 031303 (2003)[astro-ph/0208048];
 R. Guedens, D. Clancy  and A. R. Liddle,
\textit{Primordial black holes in braneworld cosmologies: Accretion after formation},  
Phys. Rev. D \textbf{66}, 083509 (2002)[astro-ph/0208299] ; 
A. S. Majumdar and N. Mukherjee, 
\textit{ Braneworld black holes in cosmology and astrophysics}, 
Int. J. Mod. Phys. D \textbf{14}, 1095 (2005)[astro-ph/0503473].

\bibitem{mgs}
A. S. Majumdar, D. Gangopadhyay and L. P. Singh, 
\textit{ Evolution of primordial black holes in Jordan-Brans-Dicke cosmology}, 
Mon. Not. R. Astron. Soc. \textbf{385}, 1467 (2008) [arXiv:0709.3193].

\bibitem{hawking}
 S. W. Hawking, 
\textit{ Black holes in the Brans-Dicke theory of gravitation}, 
Comm. Math. Phys. \textbf{25}, 167 (1972).

\bibitem{bc}
 J. D. Barrow and B. J. Carr, 
\textit{ Formation and evaporation of primordial black holes in scalar - tensor gravity theories}, 
Phys. Rev. D \textbf{54}, 3920 (1996).


\bibitem{nsm}
B. Nayak, L. P. Singh and A. S. Majumdar,
\textit{ Effect of accretion on primordial black holes in Brans-Dicke theory}, 
Phys. Rev. D \textbf{80}, 023529 (2009)[arXiv:0902.4553].


\bibitem{sbs}
Ya. B. Zel'dovich and I. D. Novikov,
\textit{The hypothesis of cores retarded during expansion and the hot cosmological model}, 
Sov. Astron. \textbf{10}, 602 (1967);
B. J. Carr and S. W. Hawking, 
\textit{ Black holes in the early Universe}, 
Mon. Not. R. Astron. Soc. \textbf{168}, 399 (1974).

\bibitem{cgim}
C. E. Fichtel et al.,
\textit{High-energy gamma-ray results from the second small astronomy satellite}, 
Astrophys. J. \textbf{198}, 163 (1975);
G. F. Chapline, 
\textit{Cosmological effects of primordial black holes}, 
Nature \textbf{253}, 251 (1975).

\bibitem{bba}
B. J. Carr and J. E. Lidsey, 
\textit{Primordial black holes and generalized constraints on chaotic inflation},
Phys. Rev. D. \textbf{48}, 543 (1993); 
B. J. Carr, J. H. Gilbert and J. E. Lidsey, 
\textit{ Black hole relics and inflation: Limits on blue perturbation spectra}, 
Phys. Rev. D. \textbf{50}, 4853 (1994) [astro-ph/9405027].


\bibitem{carretal}
B. J. Carr, K. Kohri, Y. Sendouda and J. Yokoyama,  
\textit{New cosmological constraints on primordial black holes}, 
Phys. Rev. D \textbf{81}, 104019 (2010) [arXiv:0912.5297].


\bibitem{cgl}
D. Clancy, R. Guedens A. R. Liddle,
\textit{ Primordial black holes in brane world cosmologies: Astrophysical constraints}, 
Phys. Rev. D \textbf{68}, 023507 (2003) [astro-ph/0301568].


\bibitem{sns} 
Y. Sendouda, S. Nagataki and K. Sato, 
\textit{ Constraints on the mass spectrum of primordial black holes and braneworld parameters from the high-energy diffuse photon background}, 
Phys. Rev. D \textbf{68}, 103510 (2003) [astro-ph/0309170].

\bibitem{skns}
Y. Sendouda, K. Kohri, S. Nagataki and K. Sato,
\textit{Sub-GeV galactic cosmic-ray antiprotons from primordial black holes in the Randall-Sundrum braneworld}, 
Phys. Rev. D \textbf{71}, 063512 (2005) [arXiv:astro-ph/0408369].

\bibitem{za}
Ya. B. Zel'dovich and A. A. Starobinsky,
\textit{Possibility of a cold cosmological singularity in the spectrum of primordial black holes}, 
JETP Lett. \textbf{24}, 610 (1976).

\bibitem{mc}
J. H. MacGibbon and B. J. Carr, 
\textit{ Cosmic rays from primordial black holes}, 
Astrophys. J. \textbf{371}, 447 (1991).

\bibitem{cnpsz}
I. D. Novikov, A. G. Polnarev, A. A. Starobinsky and Ya. B. Zel'dovich, 
\textit{Primordial Black Holes}, 
Astron. Astrophys. \textbf{80}, 104 (1979).
  
\bibitem{gcl}
R. Guedens, D. Clancy and A. R. Liddle,  
\textit{ Primordial black holes in braneworld cosmologies: Formation, cosmological evolution and evaporation}, 
Phys. Rev. D \textbf{66}, 043513 (2002) [astro-ph/0205149].


\bibitem{page} 
D. N. Page, 
\textit{ Particle Emission Rates from a Black Hole: Massless Particles from an Uncharged, Nonrotating Hole}, 
Phys. Rev. D \textbf{13}, 198 (1976).


\bibitem{ph}
D. N. Page and S. W. Hawking, 
\textit{ Gamma rays from primordial black holes}, 
Astrophys. J. \textbf{206}, 1 (1976).

\bibitem{smr}
A. W. Strong, I. V. Moskalenko and O. Reimer, 
\textit{Diffuse galactic continuum gamma-rays: A model compatible with egret data and cosmic-ray measurement}, 
Astrophys. J. \textbf{613}, 962 (2004) [astro-ph/0406254]. 

\bibitem{macweb}
J. H. MacGibbon and B. R. Webber, 
\textit{Quark- and gluon-jet emission from primordial black holes: The instantaneous spectra}, 
Phys. Rev. D \textbf{41}, 3052 (1990). 

\bibitem{szm}
R. A. Sunyaev and Ya. B. Zel'dovich, 
\textit{ The Interaction of matter and radiation in the hot model of the universe}, 
Astrophys. Space Sci. \textbf{7}, 20 (1970) ; 
J. C. Mather et. al., 
\textit{Measurement of the cosmic microwave background spectrum by the COBE FIRAS instrument}, 
Astrophys. J. \textbf{420}, 439 (1994).

\bibitem{fixsen}
D. J. Fixsen et al., 
\textit{The Cosmic Microwave Background Spectrum from the Full COBE/FIRAS Data Set}, 
Astrophys. J. \textbf{473}, 576 (1996) [arXiv:astro-ph/9605054].

\bibitem{nasel}
P. D. Nasel'skii,  
\textit{Hydrogen recombination kinetics in the presence of low-mass primordial black holes},  
Pis'ma Astron. Zh. \textbf{4}, 209 (1978).

\bibitem{tasug}
H. Tashiro and N. Sugiyama, 
\textit{Constraints on primordial black holes by distortions of the cosmic microwave background}, 
Phys. Rev. D \textbf{78}, 023004 (2008) [arXiv:0801.3172].


\bibitem{carr}
B. J. Carr, 
\textit{Some cosmological consequences of primordial black hole evaporations}, 
Astrophys. J. \textbf{206}, 8 (1976).

\bibitem{vdn}
B. V. Vainer, O. V. Dryzhakova, and P. D. Nasel'skii, 
\textit{Primordial black holes and cosmological nucleosynthesis}, 
Pis'ma Astron. Zh. \textbf{4}, 344 (1978) [Sov. Astron. Lett. \textbf{4}, 185 (1978)]. 

\bibitem{ky}
K. Kohri and J. Yokoyama. 
\textit{Primordial black holes and primordial nucleosynthesis: Effects of hadron injection from low mass holes},  
Phys. Rev. D \textbf{61}, 023501 (2000)[arXiv:astro-ph/9908160].

\bibitem{zskc}
Ya. B. Zel'dovich, A. A. Starobinsky, M. Yu. Khlopov and V. M. Chechekin, 
\textit{Primordial black holes and the deuterium problem}, 
Pis'ma Astron. Zh. \textbf{3}, 208 (1977) [Sov. Astron. Lett. \textbf{3}, 110 (1977)]. 

\bibitem{lindley}
D. Lindley, 
\textit{Primordial black holes and deuterium abundance},
Mon. Not. R. Astron. Soc. \textbf{193}, 593 (1980).

\bibitem{kt}
E. W. Kolb and M. S. Turner, 
\textit{The Early Universe} (Addison-Wesley, New Work, 1990).
 
\bibitem{gl}
A. M. Green and A. R. Liddle, 
\textit{ Constraints on the density perturbation spectrum from primordial black holes}, 
Phys. Rev. D \textbf{56}, 6166 (1997) [astro-ph/9704251].


\end{thebibliography}
\end{document}